\documentclass[%
 reprint,
 amsmath,amssymb,
 aps,
]{revtex4-1}

\usepackage{todonotes}

\usepackage{graphicx}
\usepackage{dcolumn}
\usepackage{bm}

\usepackage{amssymb}
\usepackage{epsfig}
\usepackage{caption}
\usepackage{subcaption}

\usepackage{bm}
\usepackage{latexsym}
\usepackage{dcolumn}
\usepackage{amsfonts}
\usepackage{graphicx}
\usepackage{epsfig}
\usepackage{amssymb}
\usepackage{nopageno}
\usepackage{amsmath}

\def \f  {\frac}

\def \bea {\begin{eqnarray}}
\def \eea {\end{eqnarray}}
\def \be {\begin{equation}}
\def \ee {\end{equation}}
\def \barr {\begin{array}{lc}}
\def \earr {\end{array}}

\begin{document}

\preprint{APS/123-QED}

\title{Quantum Non-Barking Dogs}

\author{Sara Imari Walker}
 \altaffiliation[Also at ]{School of Earth and Space Exploration, Arizona State University, Tempe AZ 85287-1504, USA and Blue Marble Space Institute of Science, Seattle WA 98145-1561, USA}
\author{Paul C.W. Davies}%
 \email{Paul.Davies@asu.edu}
 \homepage{www.beyond.asu.edu}
\affiliation{%
Beyond Center for Funamental Concepts in Science, Arizona State University, Tempe AZ 85287-1504, USA
}%

\author{Prasant Samantray}
\affiliation{International Centre for Theoretical Sciences, IISc Campus, Bengaluru 560012, India}
\author{Yakir Aharonov}
\affiliation{%
Department of Physics, Computational Science, and Engineering, Schmid College of Science, Chapman University, Orange CA 92866 USA \\
School of Physics and Astronomy, Tel Aviv University, Tel Aviv, Israel
}%

\date{\today}

\begin{abstract}
Quantum weak measurements with states both pre- and post-selected offer a window into a hitherto neglected sector of quantum mechanics. A class of such systems involves time dependent evolution with transitions possible. In this paper we explore two very simple systems in this class. The first is a toy model representing the decay of an excited atom. The second is the tunneling of a particle through a barrier.  The post-selection criteria are chosen as follows: at the final time, the ÒatomÓ remains in its initial excited state for the first example and the particle remains behind the barrier for the second. We then ask what weak values are predicted in the physical environment of the ÒatomÓ (to which no net energy has been transferred) and in the region beyond the barrier (to which the particle has not tunneled). Previous work suggests that very large weak values might arise in these regions for long durations between pre- and post-selection times. Our calculations reveal some distinct differences between the two model systems.
\end{abstract}

\maketitle

\section{Introduction}

Given that quantum mechanics is several decades old, it is remarkable that a significant sector of the theory lay unexplored until recently. To glimpse this sector it is merely necessary to consider the expectation value of an observable $A$ in a quantum state $\Psi$, assumed for the moment to be a stationary state, and note that by inserting a complete set of states it may be decomposed as follows:
\begin{eqnarray}
\langle \Psi  \mid A \mid \Psi\rangle = \sum_i \mid \langle \Psi \mid \phi_i \rangle \mid^2  \left[ \frac{ \langle \phi_i \mid A \mid \Psi \rangle}{\langle \phi_i \mid \Psi \rangle} \right] \label{eqn:ensemble}
\end{eqnarray}
where $\phi_i$ are eigenstates of a different observable $B$ ({\it i.e.}, $B$ does not commute with $A$). The first term of the summand is recognized as the probability that, on measurement of $B$, the system will be found in state $\phi_i$. The second term (in the square braces) is called the weak value of $A$ (one value for each specific $\phi_i$). For an individual system the weak value has little meaning, but it does contain non-trivial statistical information about the world when one considers a large ensemble of identical systems, each prepared in state $\Psi$. Under those circumstances, the weak value can be shown to be precisely the mean value of an ensemble of measurements of $B$ in which the coupling between the system and the measurement device is made arbitrarily weak (and hence non-disturbing) \cite{AR2005, AV1990}. Note that weak values are not eigenvalues. Rather, they are statistical averages of weak measurements. They may take values outside the spectrum of eigenvalues \cite{ACAV}. They may not even be real numbers; the real and imaginary parts have separate physical interpretation \cite{AAV1988}. Weak values can and have been measured, and are the subject of considerable theoretical and experimental interest (see {\it e.g.}, \cite{AR2005, AR1990, S1995, FKR2007, Tollaksen2007, Davies2009, Tollaksenetal2010}).

The subject of weak values becomes of greater interest when combined with postselection of states. That is possible because, for a large enough ensemble of identical systems in identically prepared initial states (preselection), there will always be a sub-ensemble of systems which are also found, on subsequent measurement of $B$, to be in any given eigenstate state $\phi_i$. A generic expression for weak values may be written, schematically,
\begin{eqnarray}
w_A = \frac{\langle \text{out} \mid A \mid \text{in} \rangle}{\langle \text{out} \mid \text{in} \rangle}
\end{eqnarray}
for observable $A$, with the system prepared in state $\mid \text{in} \rangle$, and postselected for state $\mid \text{out} \rangle$.

In this paper we consider weak values with postselection of states for systems that are intrinsically time-dependent; that is, they evolve unitarily in time away from a stationary state, so that the time dependence is not merely the result of measurement. Under these circumstances the weak value is given by
\begin{eqnarray} \label{eq:wA}
w_A(t) = \frac{\langle \text{out} \mid U^\dagger (t - t_{\text{out}}) A U (t - t_{\text{in}}) \mid \text{in} \rangle}{\langle \text{out} \mid U^\dagger (t - t_{\text{out}}) U (t - t_{\text{in}}) \mid \text{in} \rangle} \label{wA}
\end{eqnarray}
where $U(t)$ is the unitary evolution operator for the system. We consider two examples in this paper:  the decay of an excited atom, and the tunneling of a particle through a barrier. 

\section{Decay of an Excited Atom} \label{atom}

Consider an atom prepared in an excited state at initial time $t = t_i$. Over time it will decay. The decay may be described using first order perturbation theory, which predicts that the expectation value of the projection operator onto the excited state will have the well-known time dependence $e^{-\gamma(t-t_i)}$. In the context of weak measurements, it is then possible to ask the following question. Suppose the atom is inspected at some time $t_f$, and found to have definitely decayed. What result would be obtained for a weak measurement of the projection operator made at time $t$, in the interval $t_i < t < t_f$, with the state of the atom at $t_f$ postselected to be decayed (understood in the context of measurements averaged over large ensemble of identical systems)? This problem was studied by one of us (PD) using an exactly solvable model of an initially excited two-level reference atom coupled equally to all members of a large bath of similar two-level atoms, all prepared initially in their ground states \cite{Davies2009}. The weak value of the projection operator onto the excited state of the reference atom as a function of time could then be evaluated for various choices of postselection for the bath atom states. In the special case that the postselected bath state corresponds to an excited atom with energy levels that coincide with the reference atom, the result was found to be real, and simple:
\begin{eqnarray}
w = e^{-\gamma (t - t_i)} \left[ \frac{1 - e^{-\gamma (t_f - t)}}{1- e^{-\gamma (t_f - t_i)}} \right]
\end{eqnarray}	 
which reduces to the standard exponential decay law when $t_f \rightarrow \infty$. 

The same model can also be used to solve the complementary question of postselecting the atom to have definitely {\it not decayed} at time $t_f$, and that is the case we wish to address here. The weak value of the projection operator onto the excited state of the reference atom at time $t$ in that case is $w = 1$. However, our interest here lies with the bath atoms. It might be supposed that if the reference atom, henceforth labeled $0$, is both pre-and postselected to be in its excited state, and that the weak value for atom $0$ to be excited remains unchanged at $w = 1$, then weak measurements of the state of the bath in the interval $t_i < t < t_f$ would inevitably yield values corresponding to the ground states of every bath atom, since no energy will have flowed from the excited atom $0$ to the bath at the conclusion of the interval $\left[ t_i, t_f \right]$. Intriguingly, this supposition is incorrect. 

To demonstrate this, we calculate the weak values of projection operators onto the excited states of the bath atoms, in the case that atom $0$ is both pre- and postselected to be in its excited state. For simplicity, consider a population of $N$ bath atoms with upper energy levels equispaced and given by
\begin{eqnarray}
E_n - E_0 = n E ~~~~~ - N \leq n \leq N
\end{eqnarray}
{\it i.e.}, the excited states are distributed symmetrically about the excited state of atom $0$. Let the weak value for the projection operator onto the excited state of atom $n$ be denoted $w_n$, and the bra vector for the initial state of the total system be denoted as $(1,0,0,0, \ldots)$, the first entry corresponding to atom 0 in its excited state and the remaining entries to the $n$ bath atoms in their ground states. The projection operator $P_n$ onto the excited state of atom $n$ will then be, in this notation, a square matrix with all elements 0 except the entry for row $n$, column $n$, which will be 1. The Schr\"{o}dinger equation for this system is a set of coupled differential equations
\begin{eqnarray}
\dot{a}_0 &=& -i\sum_n Ha_ne^{-in\Delta Et} \nonumber\\
\dot{a}_n &=& -iH a_0 e^{in\Delta Et} \label{Eq_35}
  \end{eqnarray}
where $a_n$ is the probability amplitude that the atom labeled by $n$ is in the excited state. We set $\hbar = 1$ for convenience, and choose $H$ to be real for simplicity. In the limit that $N\to \infty,  \Delta E\to 0$, $H\to 0$, and $\frac{H^2\pi}{\Delta E}\to \gamma$ (where $\gamma$ is defined as the decay constant), the above set of equations can be solved exactly using Laplace transforms. The evolution operator (which in this case is a ${2N+1} \times {2N+1}$ matrix) can be written as:
\bea
U_{00}(t) &=& e^{-\gamma |t| - iE_0 t} \nonumber \\
U_{n0}(t) &=& H e^{-iE_n t} \frac{e^{-\gamma |t| - iE_n t - 1}}{\gamma - i n \Delta E}  \nonumber \\
U_{0n}(t) &=& H e^{-iE_n t} \frac{e^{-\gamma |t| - iE_n t - 1}}{\gamma + i n \Delta E}~. \label{evoleqs}
\eea
\noindent The elements $U_{nm}$ are not required for what follows. It may be readily verified that the above operator satisfies the unitarity constraint $UU^{\dag} = 1$ (for the elements given), and the evolution condition $U(t_f - t)U(t - t_i) = U(t_f - t_i)$. The weak values of interest are given by
\be
w_n = \frac{[1,0,0,0,..]^T U(t_f - t)P_n U(t - t_i)[1,0,0,0..]}{[1,0,0,0,..]^T U(t_f - t_i)[1,0,0,0..]}
\ee
where $[1,0,0,0,..]^T$ is the transpose of the column vector $[1,0,0,0,..]$, and use has been made of the relation $U^{\dag} (t - t_f) = U(t_f - t)$. The matrix multiplications are straightforward, and using Eq. (\ref{evoleqs}) we find, in the aforementioned limits, 
\bea
w_n &=&  \left(\frac{H^2}{\gamma^2 + n^2 \Delta E}\right)  e^{\gamma(t_f - t_i)} \Bigl( e^{-\gamma(t_f - t_i)} + e^{-in\Delta E(t_f - t_i)}  \nonumber \\ 
&& -  e^{-\gamma(t - t_i) - in\Delta E(t_f - t) } - e^{-\gamma(t_f - t) - in\Delta E(t-t_i)} ] \Bigr)~. \label{eqn:wn}
\eea
Eq. (\ref{eqn:wn}) gives the value that the $n$th atom, weakly measured, is in an excited state, subject to our specification of pre- and postselected states. Note that $w$ vanishes (by construction) at $t_f$ and $t_i$. Inspection of Eq. (\ref{eqn:wn}) reveals that the weak values of individual bath atoms are generally non-zero in the interval $[t_i,t_f]$. However, it is readily verified, in the limit $\Delta E \rightarrow 0$ where the sums may be performed explicitly, that
\begin{eqnarray} \label{eqn:sumw}
\sum_n w_n = 0
\end{eqnarray}
as it must, by unitarity.  Thus, on average, the bath atoms are undisturbed during the interval $[t_i,t_f]$. This can occur only if some $w_n$ are negative. The fact that an operator (the projection operator onto a bath atom's excited state) can be negative seems counter-intuitive to those used to thinking of quantum states in terms of eigenvalues, but it is well-established that such ``weird'' weak values are commonplace for various postselections \cite{AR2005, ACAV}.

Another distinctive property of $w_n$ is that it increases exponentially with time. This is seen most strikingly in the case that we again choose $n = 0$, {\it i.e.}, the bath atoms have the same energy levels as the reference atom, when the expressions are all real:
\begin{eqnarray}
w_0 = \frac{H^2}{\gamma^2 (1+ e^{\gamma (t_f - t_i)} - e^{\gamma(t_f - t)} - e^{\gamma(t-t_i)})}
\end{eqnarray}
The interpretation of this result is as follows. The probability of finding the chosen postselected state -- atom $0$ undecayed -- becomes exponentially small at times in excess of the normal half-life of the atom. That is, an exponentially smaller fraction of $N$ in the ensemble of identical systems will be found on measurement at time $t_f$ to be in the state with atom $0$ still excited. Thus $U(t_f - t_i)\mid \text{i} \rangle$, being the initial state unitarily evolved to the final time $t_f$, will be almost orthogonal to the postselected final state $\mid \text{f} \rangle$:
\begin{eqnarray} \label{eqn:bound}
\langle \text{f} \mid U (t_f - t_{i}) \mid \text{i} && \rangle = \nonumber \\
&& \langle \text{f} \mid U^\dagger (t - t_{f}) U(t - t_{i}) \mid \text{i} \rangle \ll 1
\end{eqnarray}
Inspection of Eq. (\ref{wA}) then indicates that $w$ will be very large (positive or negative). The longer the interval $t_{f} - t_{i}$ the smaller Eq. (\ref{eqn:bound}) becomes, and the larger the values of $w_n$ become, exponentially. Nevertheless, because $w_n$ can be both positive and negative according to the value of $n$, Eq. (\ref{eqn:sumw}) still applies. 

What can we conclude from this calculation? The result demonstrates that if an excited atom is found after a period of time to have not decayed, this does not mean the electromagnetic field in the vicinity of the atom is undisturbed. The use of weak measurements can reveal activity in the field. This activity will average to zero, but individual weak values of, say, the energy density of the electromagnetic field, will be non-zero and will in fact grow exponentially large (positive and negative) as the postselection time becomes much longer than the half-life of the excited state. Thus, just as the dog that didn't bark in Arthur Conan Doyle's story \textit{Silver Blaze} gave Sherlock Holmes meaningful information about the dog's non-canine environment, so too the atom that doesn't decay gives measurable information about physical changes in its environment (Readers of a later generation may prefer a Rolling Stone's analogy: ``I hear the telephone that hasn't rung."). 

We now turn to a second example that demonstrates the same phenomenon, but with some importantly different features.

\section{Particle tunneling through a barrier}

\noindent As a second example, let us consider the quantum tunneling of a particle trapped inside a potential well. We model the system by the potential:
\bea
V(x) &=& \f{\hbar^2 \kappa}{\mu}\delta(x) ;~~x > -2L \nonumber\\
&=& \infty~~;~~x \leq -2L  \label{QMPotential}
\eea
where $\kappa>0$, {\it i.e.}, we model the system with a potential barrier centered at $x=0$. The problem is set up as follows. A particle, modeled by a  wave packet, is moving to the right and is prepared such that at time $t = 0$ it is positioned at the center of the well at $x = -L$. Eventually it encounters the delta function potential and is partially reflected and partially transmitted. The reflected component moves leftward until it encounters the infinite wall at $x = -2L$, from which it is totally reflected, to repeat its journey in the direction of the delta function potential, from which a second reflection and transmission process occurs. This back and forth motion is repeated a large number of times. Each reflection at the delta function diminishes the amplitude of the reflected component and creates one additional transmitted packet, also with diminished amplitude. If we wait for a sufficiently long time, we will find that
the wave packet has almost completely tunneled through the potential. As with the previous case of the atom that did not decay, utilizing weak-measurements permits us to probe interesting features of the system if the particle, at some later time $T$, is found to {\it not have tunneled} through the barrier. We are interested in weak values at an intermediate time $0<t<T$ in the region outside the well, subject to the postselection that the particle is still confined to the well at time $T$. \\

\noindent For simplicity, we model the trapped particle by the Gaussian wave packet
\bea
&\Phi(x,t)& = \f{b}{\sqrt{b^2 + \f{i\hbar t}{\mu}}} e^{ ik_0 \left(x - x_0 - \f{vt}{2}\right) - \f{(x - x_0 - vt)^2}{2(b^2 + \f{i\hbar t}{m})} }
\eea
where $\mu$ is the particle mass, and $x_0$ is the position of maximal amplitude $|\Phi(x,t)|^2$ at $t=0$. To avoid the complications associated with spreading of the wave packet, we assume the following conditions:

\begin{enumerate}
\item the mass $\mu$ is very large such that $b^2 \gg \f{\hbar t}{\mu}$ in the range of time that we are interested in,
\item  the initial width $b \ll L$, and
\item the dominant wave number $k_0$ is very large such that $v = \f{\hbar k_0}{\mu}$ is finite.
\end{enumerate}

\noindent With the above conditions, the wave packet can be approximated as
\be
 \Phi(x,t) = e^{ik_0 \left(x - x_0 - \f{vt}{2}\right) - \f{(x - x_0 - vt)^2}{2b^2}}  \label{approxwavepacket}
\ee
which suffices for the purposes of our calculation. Preselecting our wave-packet to be located at $x = -L$ at $t=0$ moving to the right yields:
\be
\Phi_{{pre}}(x,0) = e^{ik_0(x + L) - \f{(x + L)^2}{2b^2}}~. \label{preselection}
\ee 
Likewise, we postselect the state at a later time $T$ to be again located at $x = -L$ and moving to the left:
\be
\Phi_{{post}}(x,T) = e^{-ik_0(x + L) - \f{(x + L)^2}{2b^2}} \label{postselection} ~.
\ee

We have chosen the postselected state to be the same as the preselected state, except for the fact that the postselected wave is traveling in the opposite direction. These choices of pre- and postselections are chosen for convenience without any loss of generality. In order to compute the weak value, one requires an exact time-dependent solution for the potential in Eq. (\ref{QMPotential}), which for an incident Gaussian wave packet is intractable. However, one can derive the reflection ($\rho$) and transmission ($\tau$) coefficients for the time-independent Schr\"odinger equation in the presence of a delta function barrier alone (with no reflecting wall)\cite{CL2008,AD2004,DN2002,Yearsley2008}, which yields:
\bea
\rho = \f{-i\kappa}{k_0 + i\kappa}~, ~~~
\tau = \f{k_0}{k_0 + i\kappa}~.
\eea

\noindent Therefore, to approximate the time-dependent solution without introducing much error, it can be safely assumed that every time a narrow Gaussian wave packet is incident upon a delta barrier, the reflected component (assumed to be a Gaussian) has its amplitude reduced by a factor $\rho$. The rest of the wave packet is transmitted with a factor $\tau$ times the amplitude of the incident wave packet. \\

To add the effects of an infinite wall to the existing delta barrier is non-trivial. For our purpose, it suffices to assume that the wall is at a large distance from the barrier, {\it i.e.}, $\f{L}{b} \gg 1$ and acts as a mirror. Also $b^2 \gg \f{\hbar t}{m}$, and the typical time scale will be given by $t \sim L/v$ from dimensional grounds. However, since $mv = \hbar k_0$, we require that
\be 
bk_0 \gg \f{L}{b} \gg 1 ~. \label{Necessary_condition} 
\ee

\noindent With these approximations, we can construct the forward time evolution of the preselected wave packet (\ref{preselection}) as
\bea
\Phi_{{pre}}(x,t) &=& \rho^N \Bigl(e^{ik_0\left(x + (4N + 1)L -\f{vt}{2} \right) - \f{\left(x + (4N + 1)L - vt \right)^2}{2b^2}} \nonumber \\
&-& e^{-ik_0\left(x - (4N - 3)L + \f{vt}{2} \right) - \f{\left(x - (4N - 3)L + vt \right)^2}{2b^2}} \Bigr) \label{fwd_left}
\eea
for the $Nth$ reflected component in the region $-2L<x<0$, and
\bea
\Phi_{{pre}}(x,t) &=& \sum^{N}_{n=1} \tau\rho^{n-1} \nonumber \\
&& \times e^{ik_0\left(x + (4n - 3)L -\f{vt}{2} \right) - \f{\left(x + (4n - 3)L - vt \right)^2}{2b^2}}\label{fwd}
\eea
for the transmitted component(s) in the region $0<x<\infty $, where $N$ is the total number of interactions with the delta barrier for the forward evolving wave-packet. As required, the solution satisfies the boundary condition at the wall $\Phi(-2L,t)=0$. It can be seen that with increasing time, the amplitude of the oscillating wave packet decreases by successive powers of the factor $\rho$. A succession of $N$ such interactions with the barrier creates a train of $N$ transmitted wave packets in the region $x>0$ as intuitively expected. 

Following similar reasoning, the backward time evolution of the postselected wave packet can be expressed as
\bea
\Phi_{{post}}(x,t) &=& \rho^M \Bigl( e^{-ik_0\left(x + (4M + 1)L + \f{v(t - T)}{2} \right) - \f{\left(x + (4M + 1)L + v(t-T) \right)^2}{2b^2}}\nonumber \\
&-& e^{ik_0\left(x - (4M - 3)L - \f{v(t - T)}{2} \right) - \f{\left(x - (4M - 3)L - v(t-T) \right)^2}{2b^2}} \Bigr) \label{Bwd_left}\nonumber \\
\eea
for the $M$th reflected component in the region $-2L<x<0$, and 
\bea
\Phi_{{post}}(x,t) &=& \sum^{M}_{m=1} \tau\rho^{m-1} \nonumber \\
&& \times e^{-ik_0\left(x + (4m - 3)L + \f{v(t - T)}{2} \right) - \f{\left(x + (4m - 3)L + v(t-T) \right)^2}{2b^2}}\label{Bwd}
\eea
for the transmitted component(s) in the region $0<x<\infty$, where $M$ denotes the total number of interactions with the delta barrier for the backward evolving wave-packet, and $T = t_{out} - t_{in}$. Recall that weak values at time $t$ for time-dependent systems are computed at a time $t$ by evolving the preselected states forward in time from $t_{in}$ to $t$ and evolving postselected states backward in time from $t_{out}$ to $t$ (see Eq. \ref{wA}). 

\begin{figure}[hbtp] 
\includegraphics[width=3in]{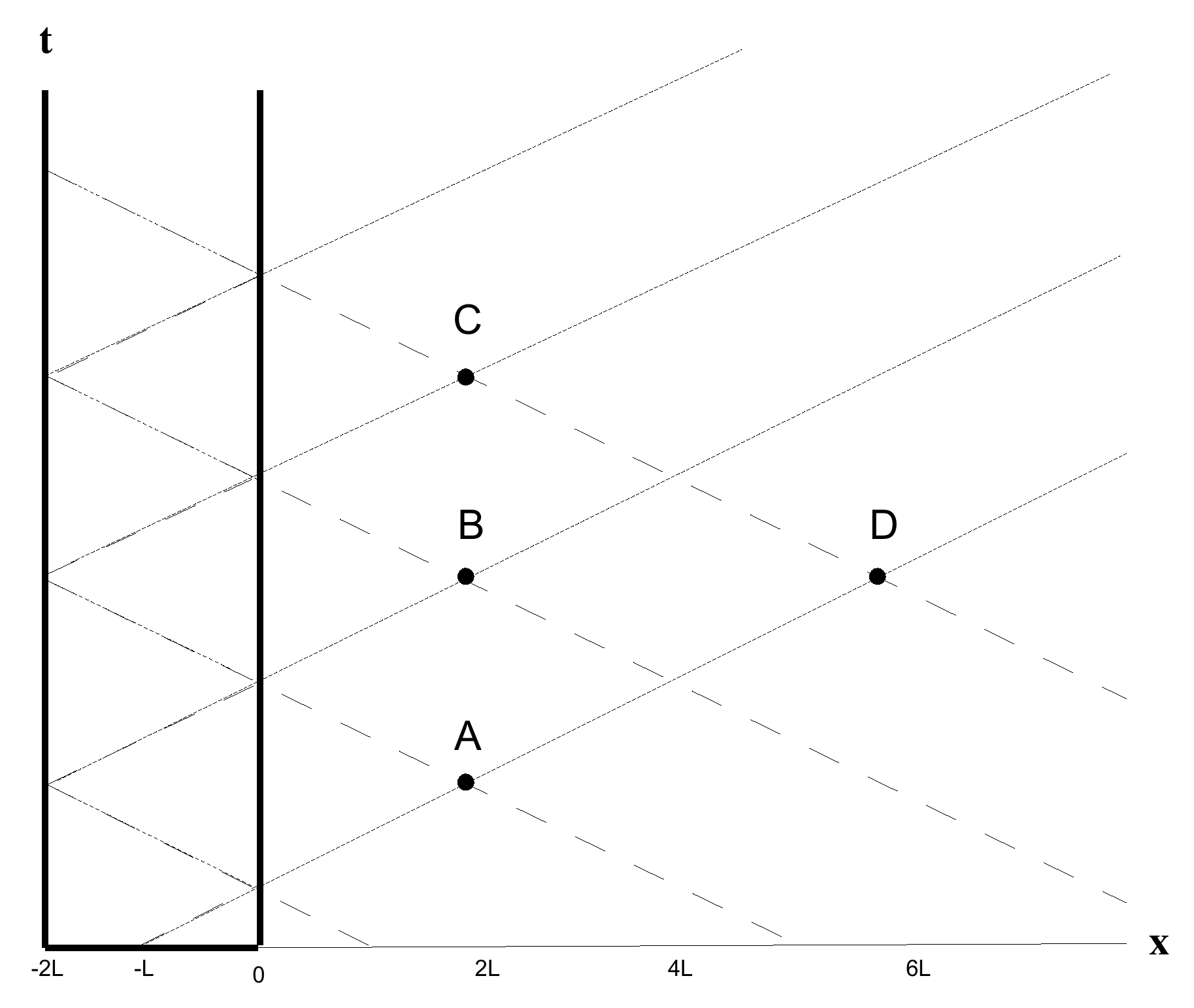}
\caption{Space-time diagram showing all sweet spots outside the well for postselected times $T = 6L/v$ (point $A$ only), and $T=14L/v$ (points $A$, $B$, $C$, and $D$). The dotted line follows the space-time trajectory of the reflected and transmitted components of the preselected wave packet, while the dashed line follows that of the postselected packet. The impenetrable wall is located at $x= -2L$ and the delta function barrier at $x=0$.}  \label{space-time}
\end{figure}

The weak value at position $x_k$ may be calculated using the projection operator 
\begin{eqnarray}
P =  \mid x-x_k  \rangle \langle x-x_k \mid
\end{eqnarray}
where $x_k > 0$ corresponds to a projection taken outside of the well, and $x_k < 0$ to a projection taken inside the well. Using this projection operator, the weak value becomes:
\begin{eqnarray} \label{wvcalc}
w(x_k, t) = \frac{\Phi_{{post}}^*(x_k,t)\Phi_{{pre}}(x_k,t)}{\int \Phi_{{post}}^*(x,t)\Phi_{{pre}}(x,t) dx}~.
\end{eqnarray}
It is easy to see that integrating this expression over all space yields,
\begin{eqnarray} \label{eqn:allspace}
\int w ~dx_k = 1~
\end{eqnarray}
which provides a useful check for the accuracy of the weak value calculations in the discussion below.

\subsection{The general solution} \label{sec:symm}

Substantial weak values will arise at space-time regions where the overlap between the forward and backward evolving wave functions is significant. Inspection of Eqs. (\ref{fwd}) and (\ref{Bwd}) reveals that the transmitted components of these forward and backward evolving wave-packets overlap at certain specific points, or ``sweet spots'' in the region $ x > 0$. Sweet spots in the space-time diagram are denoted by their coordinates $(n,m)$, with $n$ and $m$ respectively tracking the number of interactions of the pre- and postselected transmitted wave packets with the delta barrier at the time of transmission ({\it i.e.}, the contributions to the sums in Eqs. (\ref{fwd}) and (\ref{Bwd})). Fig. \ref{space-time} shows the space-time diagram of these sweet spots for the specific case of postselected times $T = 6 L/v$ and $T = 14 L/v$. Similar space-time diagrams may be drawn for any postselected time $T$ satisfying the condition
\begin{eqnarray}
 T = (4i + 2)L/v \label{Ttimes}
\end{eqnarray}
where $i=1,2,3,4\ldots$~ is an integer. The resulting spacetime diagram is symmetric, and the calculations simplify considerably without compromising important qualitative features. We therefore use postselected times consistent with Eq. (\ref{Ttimes}) throughout the remainder of this paper.

From Eq. (\ref{wvcalc}), the general solution for the weak value at any sweet spot $(n, m)$ with post-selected time consistent with Eq. (\ref{Ttimes}) is:

\begin{widetext}
\begin{eqnarray} \label{eq:full_sol}
w(n, m, N, M) &=& \frac{k_0^2 (k_0 + 2 i \kappa ) \left(\frac{\kappa }{i k_0-\kappa }\right)^{n+N+m-M}}{\sqrt{\pi } b \kappa ^2 \left(-k_0+\left(e^{b^2 k_0^2} (k_0 + 2 i \kappa ) - 2 i \kappa \right) \left(\frac{\kappa }{i k_0-\kappa }\right)^{2 N}\right)}\nonumber \\
&\times & e^\frac{\left(b^2 k_0+L (-2 i+(2+2 i)n-(2+2 i) N-(2-2 i) m+(2-2 i) M)+i x\right) \left(b^2 k_0+i (L (-2+(2+2 i) n-(2+2 i) N+(2-2 i) m-(2-2 i) M)+x)\right)}{b^2}\nonumber \\
\end{eqnarray}
\end{widetext}

\noindent The above expression is valid for any $N < M$, subject to the constraints $n \leq N$ and $m \leq M$. Thus, Eq. (\ref{eq:full_sol}) is valid for any sweet spot in the lower half of the space-time diagram where the number of forward reflections of the preselected packet is less than the number of backward reflections of the postselected packet (by symmetry an analogous expression applies for the upper half plane with $ (n, N) \leftrightarrow (m, M) $ and $ M < N $). We will first consider this result for the specific post-selection times $T = 6L/v$ and $T = 14L/v$ before returning to consider the behavior of the weak value for arbitrary post-selection time $T$.

\subsubsection{Postselection at $T = 6L/v$} \label{sec:6L}

\begin{figure}[hbtp] 
\begin{center}
\begin{subfigure}[b]{0.45\textwidth}
\includegraphics[width=2.5in]{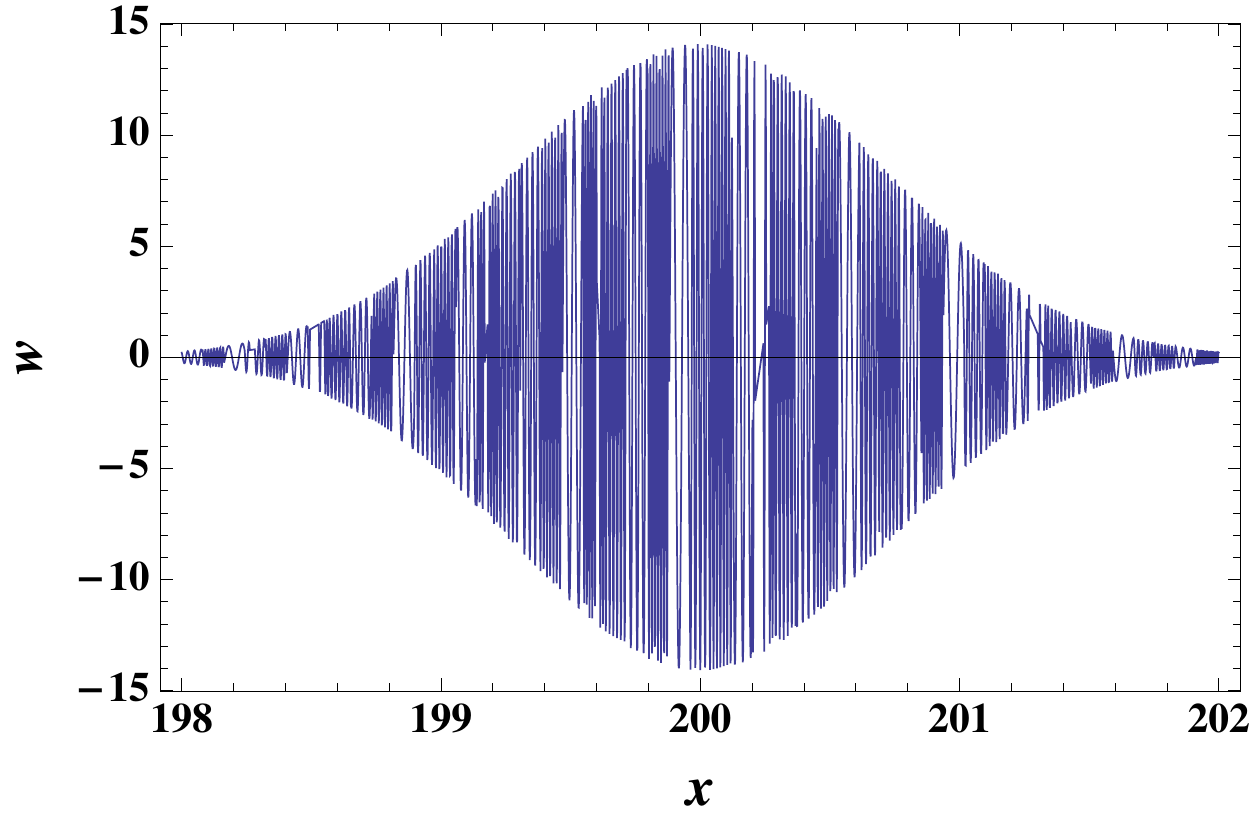}
\caption{Real Part}
\end{subfigure}
\begin{subfigure}[b]{0.45\textwidth}
\includegraphics[width=2.5in]{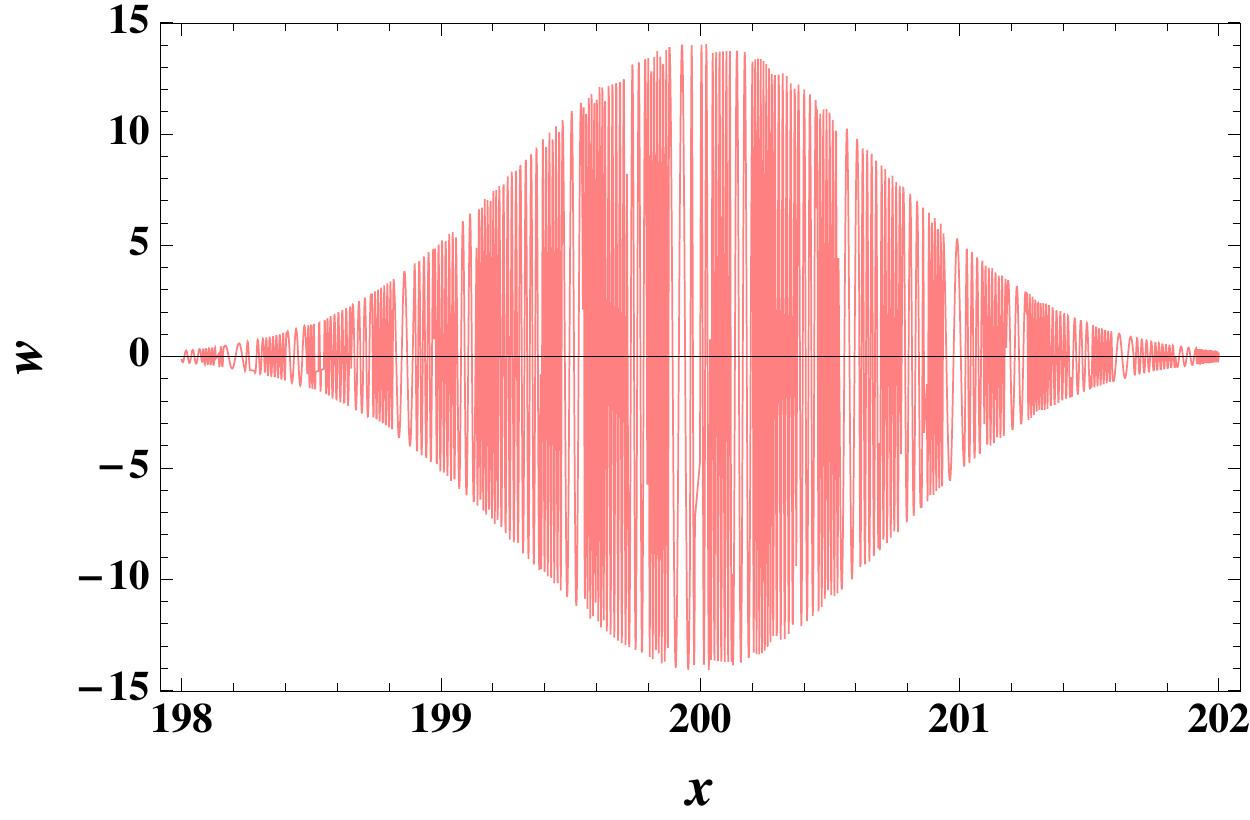}
\caption{Imaginary Part}
\end{subfigure}
\end{center}
\caption{Shown are the (A) real (blue) and (B) imaginary (pink) components of the weak value outside the well for postselection at $T=6L/v$. Constants are set as $b=1$; $\mu=1,000$;  $\kappa=1,000$; $k_0=5,000$;  and $L=100$.}   \label{fig:11out}
\end{figure}

In the atomic model discussed in Section \ref{atom}, the condition in Eq. (\ref{eqn:sumw}) ensured that the weak values in the set of bath atoms averaged to zero, consistent with the reference atom being in its excited state at the postselected time. Similarly in the present model, we expect the weak values outside the well, though individually non-zero (and potentially very large), also to average to zero. This is indeed the case, as may be shown explicitly in the simplest case where the wave-packet interacts only once with the delta function barrier before postselection at $T = 6L/v$. For this case, $N = M = 1$ and the weak measurement is taken at $t = 3L/v$ (corresponding to point $A$ in Fig. \ref{space-time}). Using Eq. (\ref{eq:full_sol}) the weak value is
\begin{eqnarray}
w(x, N &=& 1, T=6L/v) = \nonumber \\
&& \frac{ k_0^2 e^{\frac{\left(b^2 k_0 + i (x-2L)\right)^2}{b^2}}}{\sqrt{\pi } b \left(k_0^2 + \left(-1+e^{b^2 k_0^2}\right) \kappa ^2\right)}~, ~~ x > 0~. \nonumber \\ \label{6LWV}
\end{eqnarray}
This solution oscillates rapidly between positive and negative values as a function of $x$ for both the real and imaginary parts, with  decreasing magnitude as one moves away from the central sweet spot location at $x = 2L$ as shown in Fig. \ref{fig:11out}. 

Using Eqs. (\ref{fwd_left}), (\ref{fwd}), (\ref{Bwd_left}) and (\ref{Bwd}) we obtain:
\begin{eqnarray}
w(x, N &=& 1, T=6L/v) = \nonumber \\
&-&\frac{ e^{\frac{\left(b^2 k_0-i (2 L+x)\right)^2}{b^2}} \left(-1+e^{2 i k_0 (2 L+x)}\right)^2 \kappa ^2}{\sqrt{\pi } b\left(k_0^2+\left(-1+e^{b^2 k_0^2}\right) \kappa ^2\right)}~, ~~~ x < 0 \nonumber \\
\label{eqn:in}
\end{eqnarray}\\
for the reflected component inside the potential well. It is readily verified that $\int_{-2L}^{\infty} w(x)_{{inside}} dx + \int_{0}^{\infty} w(x)_{{outside}} dx = 1$, as expected from Eq. (\ref{eqn:allspace}).

\subsubsection{Postselection at $T = 14L/v$}

\begin{figure}[hbtp]  
\begin{center}
\begin{subfigure}[b]{0.45\textwidth}
\includegraphics[width=2.5in]{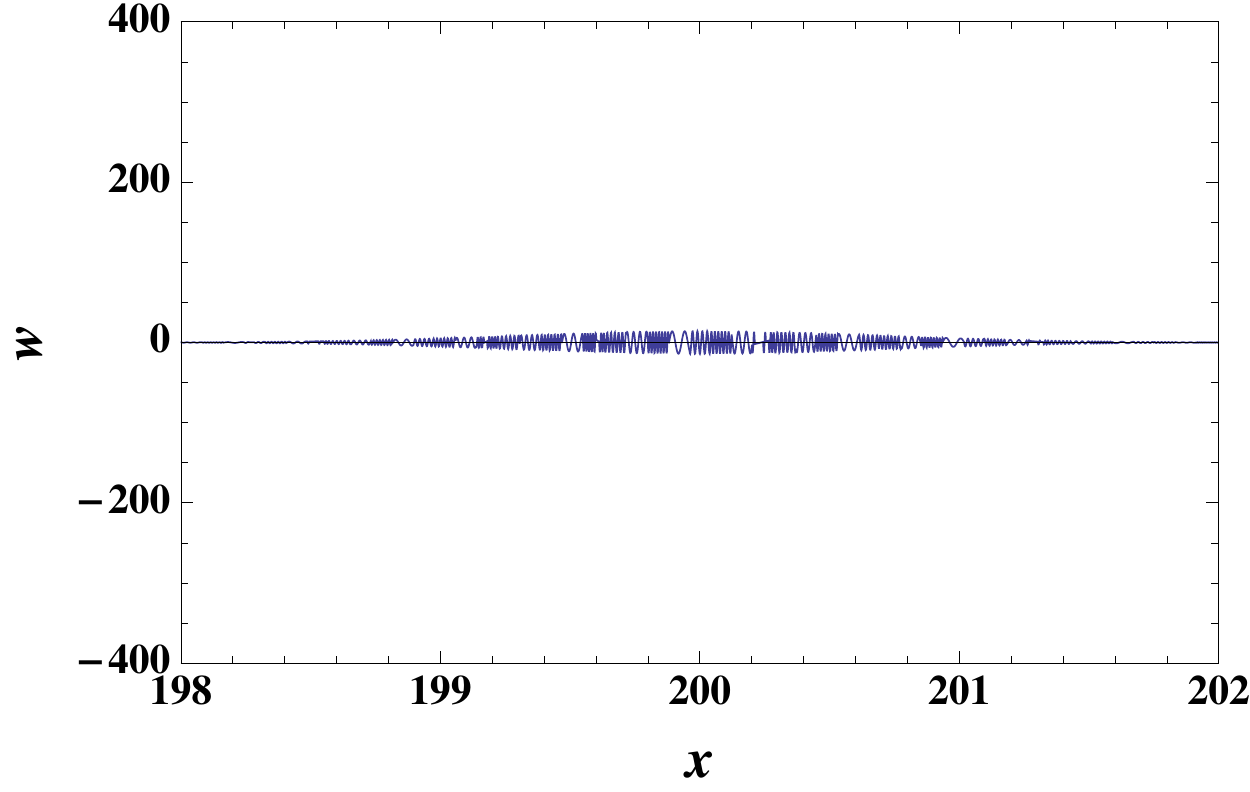}
\caption{$N = M = 2$}
\end{subfigure}
\begin{subfigure}[b]{0.45\textwidth}
\includegraphics[width=2.5in]{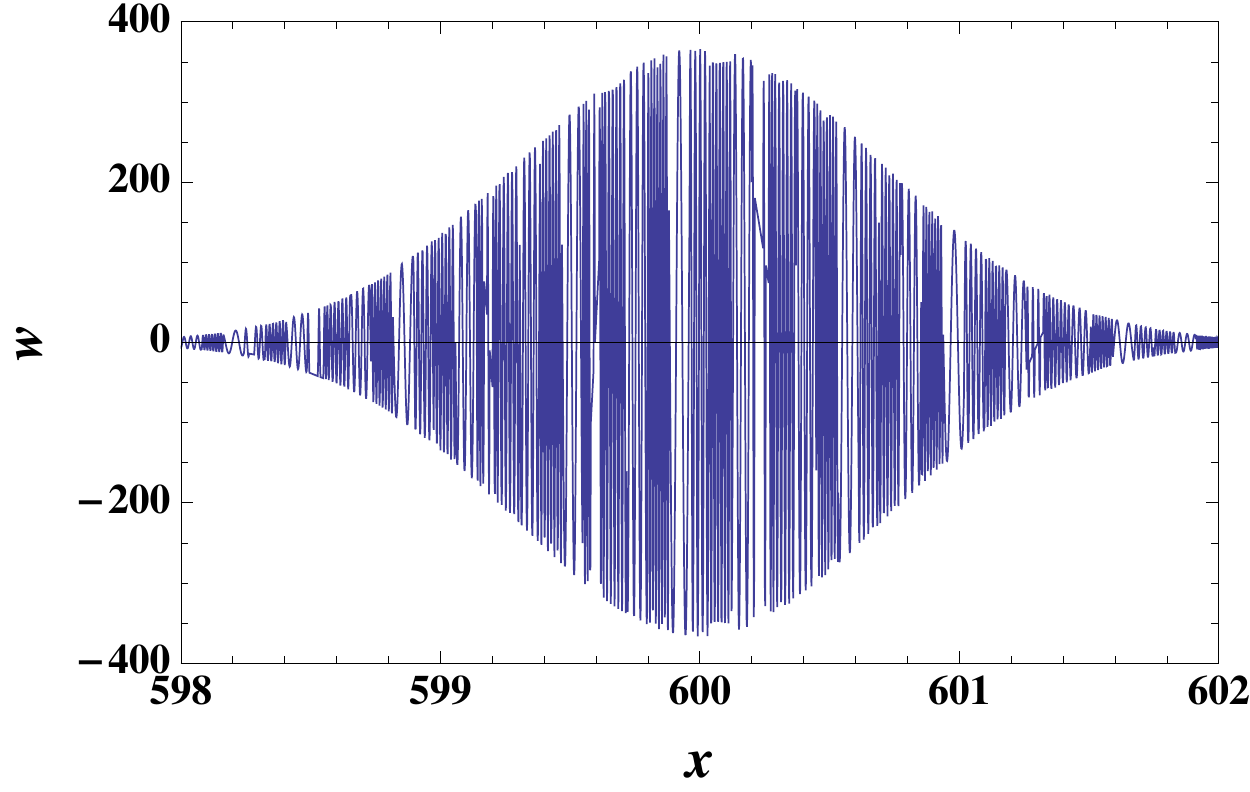}
\caption{$N = M = 1$}
\end{subfigure}
\end{center}
\caption{Shown are the  weak values  for postselection at $T=14L/v$ calculated for (A) $N = M = 2$, centered at $x = 2L$ corresponding to point $B$ in Fig. \ref{space-time} and (B) $N = M = 1$, centered at $x = 6L$, corresponding to point $D$ in Fig. \ref{space-time}. Constants are set as $b=1$; $\mu=1,000$;  $\kappa=1,000$; $k_0=5,000$;  and $L=100$.}   \label{fig:14Lv}
\end{figure}

A more interesting case is presented for larger postselected times $T$, where the number of interactions with the delta barrier is $N > 1$. Consider for example, the case for postselected time $T = 14L/v$  shown in Fig. \ref{space-time}. Eq. (\ref{eq:full_sol}) may be used to calculate the weak value at the specific space-time points B ($n = m = 2$) and D ($n = m = 1$), where $t = 7L/v$ and $N = M = 2$. The weak values are: 
\begin{eqnarray}
w(x, n &=& 2, T = 14L/v)_B= \nonumber \\
&& -\frac{e^{\frac{\left(b^2 k_0-2 i L+i x\right)^2}{b^2}} k_0^2 \kappa ^2}{\sqrt{\pi } b \left(k_0^4 + 2 i k_0^3 \kappa -2 k_0^2 \kappa ^2-\left(-1+e^{b^2 k_0^2}\right) \kappa ^4\right)} \nonumber \\ \label{14Ln2}
\end{eqnarray}
and 
\begin{eqnarray}
w(x, n &=& 1, T = 14L/v)_D = \nonumber \\
&& \frac{ e^{\frac{\left(b^2 k_0-6 i L+i x\right)^2}{b^2}} k_0^2 (k_0 + i \kappa )^2}{\sqrt{\pi } b \left(k_0^4 + 2 i k_0^3 \kappa -2 k_0^2 \kappa ^2-\left(-1+e^{b^2 k_0^2}\right) \kappa ^4\right)} \nonumber \\ \label{14Ln1}
\end{eqnarray}
for points $B$ and $D$, respectively. These expressions are plotted in Fig. \ref{fig:14Lv} and show the characteristic oscillatory behavior.  Comparing Fig. \ref{fig:14Lv} (A) and (B) demonstrates that the weak value increases as one moves away from the delta function barrier (as we discuss in the subsequent section, the ratio of Eqs. (\ref{14Ln2}) and (\ref{14Ln1}) taken at the peaks at $x = 2L$ and $x = 6L$, respectively, is consistent with Eq. (\ref{ratioNM})).

\subsubsection{Behavior of the weak value for arbitrary postselection time $T$ } \label{sec:symm}

To gain insight into the behavior of the weak value as a function of the time of post-selection, we evaluate Eq. (\ref{eq:full_sol}) for the interior set of points where $n = N$ and $m = M$ ({\it i.e.}, the time series of sweet spots centered at $x = 2L$). Thus, the results of Section \ref{sec:6L} represent a special case of those presented here for $N = M = 1$. For general $N$ and $M$:
\begin{eqnarray} \label{eq:wNM}
w(&N&, M, N, M) = \nonumber \\
&& \frac{e^{\frac{\left(b^2 k_0-2 i L+i x\right)^2}{b^2}} k_0^2 (k_0+2 i \kappa ) \left(\frac{\kappa }{i k_0-\kappa }\right)^{2 N}}{\sqrt{\pi } b \kappa ^2 \left(-k_0+\left(e^{b^2 k_0^2} (k_0+2 i \kappa) - 2 i \kappa \right) \left(\frac{\kappa }{i k_0-\kappa }\right)^{2 N}\right)} \nonumber \\ 
\end{eqnarray}
This solution is {\it independent} of number of reflections of the backward evolving wave packet $M$, and thus independent of the postselection time $T$. For any arbitrarily chosen postselected time (consistent with Eq. (\ref{Ttimes})), we arrive at the curious result that the weak value at fixed space-time point $x$ and $t$ does not depend on how far into the future we choose the postselected state. Thus, for example, a weak measurement centered at point $A$ in Fig. \ref{space-time} (with $N = 1$) taken at $t = 3L/v_0$ will yield a weak value given by  Eq. (\ref{6LWV}) for {\it any} postselected time $T$ consistent with Eq. (\ref{Ttimes}). Likewise due to the symmetry of the problem, the same weak value will be observed for a measurement again centered at $x = 2L$ with $t = T - 3L/v_0$  ({\it i.e.}, with $M = 1$) for any $T$.

Therefore, the weak value can only increase as we move away from the delta-function barrier ({\it e.g.} for fixed postselection time $T$). For fixed $N$ and $M$ as one moves away from the delta barrier, comparing the weak value at any two neighboring sweet spots yields: 
\begin{eqnarray}
\frac{w(n, m, N, M)}{w(n-1, m-1, N, M)}=\frac{\kappa ^2}{(i k_0+\kappa )^2} \label{ratioNM}
\end{eqnarray}
for $1 < n \leq N$ and $1< m \leq M$ (this of course is not unexpected since on each successive interaction with the delta barrier the transmitted component of the wave packet will diminish in amplitude by a factor $\tau \rho$). Thus, along the central axis $N = M$ the largest weak value is located at $n = m = 1$ with an amplification factor $\left( \frac{\kappa ^2}{(i k_0+\kappa )^2} \right)^{N-1}$ over the interior most point at centered at $x = 2L$. 

\section{Discussion} 

When combined with postselection of states, quantum weak measurements can expose a hidden, measurable, and hitherto neglected, sector of (standard) quantum mechanics. In this paper we have explored one aspect of this sector, namely, time-dependent systems that undergo a transition. We discussed two idealized models: the decay of an excited atom into a large bath of unexcited atoms, and the tunneling of a particle, represented by a Gaussian wave packet, through a thin barrier. The system is prepared at time $t_i$ with the reference atom excited and the bath atoms in their ground states, and the packet concentrated behind the barrier, respectively. We chose the final states to correspond to a situation where the system had not undergone the transition, that is, at time $t_f$ the atom is determined to be excited in the first example, and the particle is found to be behind the barrier in the second example. It might be supposed that since the system at the end of the experiment is in essentially the same state as it was at the beginning, nothing much of interest could be said about the external region (bath atoms and region outside the barrier, respectively). This is, however, incorrect. Weak measurements in the interval $[t_i,t_f]$ can uncover cryptic activity there. 

The systems considered have a natural half-life for the transition to occur. The probability that it has failed to occur by time $t_f$ falls sharply when $t_f - t_i$ exceeds this half-life. Nevertheless, given a sufficiently large ensemble of identically prepared systems, there will always be a sub-ensemble that satisfies both the pre- and postselection criteria. The strength of the cryptic activity in the environs rises with time, {\it i.e.}, the lower the probability of finding a given system in the postselected state, the bigger the disturbance revealed by the weak measurements. An important feature of the weak values concerned is that they must average to zero, a condition that follows from unitarity. Thus we find weak values of opposite sign that do indeed sum to zero. Although the net disturbance in the external region is zero, individual weak measurements can be very large, and grow larger with time. The negative weak values found in these examples are strange. We consider projection operators of the bath atoms onto their excited states. Such projection operators always have eigenvalues $\geq 0$. Yet weak values routinely lie outside the spectrum of eigenvalues: many other examples may be found in the literature (see {\it e.g.}, \cite{AR2005, AV1990, ACAV, AAV1988}). Similar comments apply to our second example.

An unexpected difference emerged between our two examples. In the case of the excited atom, the bath atoms displayed weak values that rose exponentially with time: the longer the atom was ÒdetainedÓ in its excited state, the bigger the weak values grew among the bath atoms. Their exponential rise mirrors the exponential nature of the decay (or the exponentially small probability of finding members of the ensemble still un-decayed at late times). In the tunneling example, however, there was steady growth, but no exponential surge in weak values. The probability of finding the particle trapped behind the barrier decreases as a power-law in the number of reflections $N$. One critical difference between the two models, which may explain the results, is that, in the one-dimensional tunneling case, there is a finite probability that the particle will tunnel through the barrier and then tunnel back again. (A Gaussian wave packet moving to the right will still contain some plane-wave components representing left-moving particles.) This probability will be independent of how far the tunneled particle has traveled to the right. On the other hand, for an atom decaying into a large bath, the probability that decay will be followed by a transfer of energy \emph{back} from the bath to the atom approaches zero in the limit of an infinite bath.

Although our examples are idealized and simplified, we contend that they capture an important and quite general feature of time-dependent quantum systems. We make no attempt here to discuss the practicality of measuring the weak values in the systems' environment, but we are confident that realistic quantum systems that reveal this aspect can be found.

\acknowledgements
The authors wish to thank Lev Vaidman and Jeff Tollaksen for constructive conversations. This work has been supported in part by the Israel Science Foundation Grant No. 1125/10.

\end{document}